\documentclass[twocolumn]{emulateapj}
\usepackage{natbib}

\slugcomment{DRAFT: \today}

\begin{document}

\newcommand{\msun}{\ensuremath{\rm M_\odot}}
\newcommand{\msunyr}{\ensuremath{\rm M_{\odot}\;{\rm yr}^{-1}}}
\newcommand{\Ha}{\ensuremath{\rm H\alpha}}
\newcommand{\Hb}{\ensuremath{\rm H\beta}}
\newcommand{\lya}{\ensuremath{\rm Ly\alpha}}
\newcommand{\Ntwo}{[\ion{N}{2}]}
\newcommand{\kms}{\textrm{km~s}\ensuremath{^{-1}\,}}
\newcommand{\ztwo}{\ensuremath{z\sim2}}
\newcommand{\zthree}{\ensuremath{z\sim3}}
\newcommand{\feh}{\textrm{[Fe/H]}}
\newcommand{\afeh}{\textrm{[$\alpha$/Fe]}}
\newcommand{\nifeh}{\textrm{[Ni/Fe]}}
\newcommand{\othree}{\textrm{[O\,{\sc iii}]}}
\newcommand{\otwo}{\textrm{[O\,{\sc ii}]}}
\newcommand{\ntwo}{\textrm{[N\,{\sc ii}]}}

\newcommand{\sitwo}{\textrm{Si\,{\sc ii}}}
\newcommand{\oone}{\textrm{O\,{\sc i}}}
\newcommand{\ctwo}{\textrm{C\,{\sc ii}}}
\newcommand{\sifour}{\textrm{Si\,{\sc iv}}}
\newcommand{\cfour}{\textrm{C\,{\sc iv}}}
\newcommand{\fetwo}{\textrm{Fe\,{\sc ii}}}
\newcommand{\altwo}{\textrm{Al\,{\sc ii}}}
\newcommand{\hetwo}{\textrm{He\,{\sc ii}}}
\newcommand{\ciii}{\textrm{C\,{\sc iii}]}}

\newcommand{\dvis}{\ensuremath{\Delta v_{\rm IS}}}
\newcommand{\hst}{{\it HST}-WFC3}
\newcommand{\cgs}{\textrm{erg s$^{-1}$ cm$^{-2}$}}

\shortauthors{Law et al. 2015}
\shorttitle{Morphological Pair Fraction at $z\sim2$}

\title{Physical Properties of a Pilot Sample of Spectroscopic Close Pair Galaxies at $z \sim 2$}

\author{David R. Law\altaffilmark{1},  Alice E. Shapley\altaffilmark{2}, Jade Checlair\altaffilmark{3}, Charles C. Steidel\altaffilmark{4}}

\altaffiltext{1}{Space Telescope Science Institute, 3700 San Martin Drive, Baltimore, MD 21218, USA (dlaw@stsci.edu)}
\altaffiltext{2}{Department of Physics and Astronomy, University of California, Los Angeles, CA 90095, USA}
\altaffiltext{3}{Dunlap Institute for Astronomy \& Astrophysics, University of Toronto, 50 St. George Street, Toronto M5S 3H4, Ontario, Canada}
\altaffiltext{4}{California Institute of Technology, MS 249-17, Pasadena, CA 91125, USA}

\begin{abstract}

We use {\it Hubble Space Telescope} Wide-Field Camera 3 ({\it HST}/WFC3)
rest-frame optical imaging to select a pilot sample of star-forming galaxies in the redshift range $z = 2.00-2.65$
whose multi-component morphologies are consistent with expectations for major mergers.
We follow up this sample of major merger candidates 
with  Keck/NIRSPEC longslit spectroscopy obtained in excellent seeing conditions (FWHM $\sim 0.5$ arcsec) to obtain
H$\alpha$-based redshifts of each of the morphological components in order to distinguish 
spectroscopic pairs from false pairs created by projection along the line of sight.  
Of six pair candidates observed, 
companions (estimated mass ratios 5:1 and 7:1) are detected for  two galaxies down
to a $3\sigma$ limiting emission-line flux of $\sim 10^{-17}$ erg s$^{-1}$ cm$^{-2}$.
This detection rate is consistent with a $\sim 50$\% false pair fraction at such angular separations
($1-2$ arcsec), and with recent claims that
the star-formation rate (SFR) can differ by an order of magnitude between the components in such mergers.
The two  spectroscopic pairs identified
have total SFR, SFR surface densities, and stellar masses  consistent on average with the overall $z\sim2$ star forming galaxy population.

\end{abstract}


\section{INTRODUCTION}

At redshift $z\sim 2-3$ galaxies are growing rapidly and build up a large fraction of their
present-day stellar mass \citep[e.g.][]{reddy08}.  As they grow, the increased stellar mass is
thought  to stabilize these systems against gravitational instabilities resulting from their large gas fractions
\citep[e.g.,][]{kassin14, vanderwel14},
decreasing their formerly-high gas-phase velocity dispersions \citep{law07b,law09,fs09,newman13} and causing a morphological 
transformation from highly-irregular clumpy starbursts \citep[e.g.,][and references therein]{guo12,law12a,vanderwel14} to the modern-day Hubble sequence \citep[e.g.,][]{papovich05,law12b,conselice14}.

One mechanism by which such growth occurs is the conversion of massive gas reservoirs into stars.
Such star formation is observed to occur at a typical rate $\sim 30 M_{\odot}$ yr$^{-1}$ 
for rest-UV selected galaxy samples \citep[e.g.,][]{erb06,wuyts11}, although this SFR may represent only a small fraction
of the gas continually cycling into \citep[e.g.,][]{genzel08,dekel09}
and out of the galaxies \citep[e.g.,][]{steidel10} through large-scale gas flows.
Likewise, galaxies also grow through both major (mass ratio $3:1$ or lower) and minor (mass ratio $4:1$ or higher) mergers with other galaxies.
Such events
typically contribute both stars and gas, thereby building up the galactic stellar spheroid population and providing fuel for future generations of star formation.
The role of mergers and merger-induced star formation compared to in-situ star formation in building up the present day galaxy population has been the subject
of considerable debate,
with various studies claiming both that mergers are
\citep{deravel09,puech14,tasca14} and are not \citep{shapiro08,williams11,wuyts11,kaviraj13} 
major drivers of star formation and galactic stellar mass assembly since $z \sim 4$.

Significant effort  has therefore been invested both in constraining
the evolution of the merger fraction for star-forming galaxies \citep[e.g.,][]{conselice08,conselice11,lotz08,lotz11,rawat08}
and in assessing the physical effects of such mergers on the star formation properties of the galaxies \citep[e.g.,][]{law07a,law12a,lotz08,lee13}.
One method employed by such studies is to use
high-resolution imaging to quantify disturbances and irregularities in the surface brightness profile
using a variety of non-parametric indices \citep[e.g.,][]{conselice00,lotz04,law07a}.
However, it is often challenging to intepret such indices unambiguously  because 
$z\sim 2-3$ galaxies are intrinsically clumpy and irregular and
similar disturbed morphologies 
can arise both in merging systems and in isolated star forming galaxies due to internal dynamical
instabilities  \citep[e.g.,][]{bournaud09,genzel11}.

An alternative way of identifying major mergers is to look for close angular pairs  ($r \lesssim 50$ kpc, $6$ arcsec at $z\sim2-3$).
When the velocity separation between the two components in such a pair is $\lesssim 500$ \kms \citep[see discussion by][]{lin04,lotz08}
numerical simulations suggest that such systems should predominantly trace major galaxy-galaxy mergers during their first pericentric passage and
before final coalescence \citep{lotz08,lotz10}.  Indeed, when merger rates derived from such close pairs \citep[e.g.,][]{bundy09,williams11,deravel09,lopez13,tasca14}
are combined with physically motivated timescales for interaction  the overall agreement on merger rates between different studies is relatively good
\citep{lotz11}.  

One major complication faced by such efforts to constrain the merger rate (and the physical characteristics of the merging galaxies) however is
the incidence of {\it false} pairs resulting from chance angular alignments of galaxies separated by large cosmological distances.  In the absence of
kinematic information, the false morphological pair fraction can be  greater than 50\% depending on the adopted impact parameter
\citep[see, e.g., discussion by][]{patton08,quadri10,chou12,law12a} and whether or not photometric redshift selection 
techniques \citep[e.g.,][]{kartaltepe07,bundy09} can be used to help trim the list of potential companions.
Although corrections for the false-pair contribution can also be estimated statistically, such statistical corrections do not identify {\it which}
are the true physical pairs, leading to substantial uncertainty in the derived properties of major mergers as a galaxy class.

In a recent contribution \citep{law12a} we used {\it HST} imaging data to quantify 
the major merger fraction in rest-UV selected (${\cal R} \leq 25.5$)\footnote{All magnitudes are given in AB units unless otherwise noted.} 
 star forming galaxies at $z \sim 2-3$ using rest-frame optical morphology.
Consistent with  studies of similar galaxy samples in the literature
\citep[e.g.,][]{conselice11} we found that $23^{+7}_{-6}$\% of these galaxies in the redshift range $2.0 < z < 2.5$ were apparent morphological pairs
with physical separations of $\leq 16$ kpc ($\sim 2$ arcsec), and $16^{+7}_{-6}$\% were statistically likely to be genuine physical pairs.
At such small separations, the pair morphology would not be apparent in the ground-based imaging
that forms the backbone of our  star forming galaxy sample.

Here we follow up the morphological pair sample presented by \citet{law12a} with rest-frame optical spectroscopy to determine which of the apparent pairs
have similar spectroscopic redshifts suggesting
that are likely to coalesce within the next $\sim$ 500 Myr \citep[see Table 5 of][]{lotz10}, and to identify whether these spectroscopically confirmed pairs have physical
characteristics that are any different from those of the rest of the star forming galaxy population.

We give an overview of the parent galaxy sample in \S \ref{sample.sec} and
describe the Keck/NIRSPEC spectroscopic follow-up observations targeting rest-optical nebular emission lines
in \S \ref{nirspec.sec}.  In \S \ref{results.sec} we present our results for individual galaxies 
and discuss the implications of these results for the physical characteristics of genuine merging pairs in \S \ref{disc.sec}.
Throughout our analysis,we adopt a standard $\Lambda$CDM cosmology based on the seven-year Wilkinson Microwave Anisotropy
Probe (WMAP) results \citep{komatsu11} in which $H_0 = 70.4$ \kms\ Mpc$^{-1}$, $\Omega_{\rm M} = 0.272$, and $\Omega_{\Lambda} = 0.728$.

\section{Target Galaxy Sample}
\label{sample.sec}

Targets were drawn from the Keck Baryonic Structure Survey \citep[KBSS;][]{trainor12} for which
galaxies are originally identified by $U_{n} G {\cal R}$ color selection down to ${\cal R} = 25.5$
and spectroscopically confirmed to lie at $z \sim 2-3$  using Keck/LRIS rest-UV spectroscopy \citep{adelberger04,steidel04}.  
As discussed by \citet{conroy08}, these galaxies are expected to typically evolve into $\sim L^{\ast}$ systems by the present day.

Of the few thousand galaxies in KBSS, 306 lie within $\sim 1$ arcminute of the line of sight to bright background
QSOs and had rest-frame optical imaging data obtained using {\it HST}/WFC3 as a part of 
Cycle 17 program GO 11694 (PI: Law). 
The details of these observations have been described at length by 
\citet{law12a}.  In brief, we used the F160W ($\lambda_{\rm eff} = 15369$ \AA) filter to trace rest-frame optical emission from
the target galaxies; these data reach a depth of 27.9 AB for a $5\sigma$ detection within a 0.2 arcsec radius aperture
and have a typical PSF FWHM of 0.18 arcsec (corresponding to $1.5$ kpc at $z\sim2$).

We selected galaxies visually classified by \citet{law12a} as `Type II', for which the rest-optical morphology consists of two or 
more distinct nucleated sources of comparable ($<$ 10:1) $H_{160}$ magnitude
and little to no evidence for extended low surface brightness features connecting the two components 
\citep[hence reducing the likelihood of multiple
clumps within a single low surface-brightness disk, e.g.,][]{genzel11,guo12}.
This selection resulted in 56 galaxies, which we further restricted to the subset with secure redshifts in the range $z = 2.00 - 2.65$ for 
which H$\alpha$ emission lies in the $K$-band between strong atmospheric absorption bands.
Additionally, we restricted our sample to 
galaxies with pair separations greater than 0.5 arcsec (the minimum separation at which we can distinguish objects in 
seeing-limited follow-up spectroscopy),
and less than 2 arcsec (the approximate limit beyond which objects can be unambiguously separated in the ground-based photometry that forms the backbone of the KBSS sample,
and would have been classified as two separate galaxies).  We also rejected targets 
whose H$\alpha$ emission was expected to lie extremely close to a bright night sky emission line feature.
In total, seven galaxies met our combined selection criteria for the 2012A observing season, of which we were able to observe six targets
in the available time.

The morphologies of these six galaxies are shown in Figures \ref{mainfig1.fig} - \ref{mainfig4.fig}; we
label the primary, secondary, and if appropriate tertiary pieces as `1', `2', and `3' respectively.\footnote{Note that 
primary, secondary, and tertiary features are generally labelled corresponding to their proximity to the centroid of the ground-based object detections and do not necessarily
correspond with $H_{160}$ magnitude.}
Assuming that the $H_{160}$ magnitude of these pieces is a proxy for their total stellar mass, we estimate that the mass ratio of these mergers ranges from 1:1
to $\sim$ 8:1 (see Table \ref{morphs.tab}).

\begin{deluxetable*}{lccccccc}
\tablecolumns{8}
\tablewidth{500pt}
\tablecaption{Target Galaxies}
\tablehead{
\colhead{Galaxy} & \colhead{R.A.\tablenotemark{a}} & \colhead{Decl.\tablenotemark{a}} & \colhead{Secondary Distance\tablenotemark{b}} & \colhead{Tertiary Distance\tablenotemark{c}} & \colhead{$H_{160}$} & \colhead{Secondary} & \colhead{Tertiary}\\
\colhead{} & \colhead{(J2000)} & \colhead{(J2000)} & \colhead{(kpc)} & \colhead{(kpc)} & \colhead{(Primary)} & \colhead{Mass Ratio\tablenotemark{d}} & \colhead{Mass Ratio\tablenotemark{d}}
}
\startdata
Q1217-BX116 & 12:19:31.271 & +49:41:21.90 & 14 & ... & 24.56 &  1:1 & ... \\
Q1217-MD16 & 12:19:28.407 & +49:40:50.15 & 9 & ... & 23.44 & 4:1 & ... \\
Q1623-BX543 & 16:25:57.736 & +26:50:09.44 & 9 & 7 & 22.82 & 8:1 & 5:1 \\
Q1700-MD103 & 17:01:00.321 & +64:11:55.42 & 13 & ... & 22.51 & 2:1 & ... \\
Q2206-BM64 & 22:08:52.360 & -19:43:28.27 & 10 & 13 & 24.03 & 1:1 & 1:1 \\
Q2343-BX429 & 23:46:22.968 & +12:49:05.55 & 10 & ... & 24.71 & 7:1 & ... \\
\enddata
\tablenotetext{a}{Coordinates represent the approximate point midway between the components based on the HST/WFC3 imaging data.}
\tablenotetext{b}{Projected distance of secondary from primary source.}
\tablenotetext{c}{Projected distance of tertiary from primary source.}
\tablenotetext{d}{Estimated from ratio of $H_{160}$ magnitudes.}
\label{morphs.tab}
\end{deluxetable*}

As described by \citet{law12a}, $H_{160}$ magnitudes for these galaxies have been combined with ground-based optical imaging
(and in some cases ground-based $J$, $K$ infrared imaging and/or {\it Spitzer}/IRAC photometry) to produce stellar population synthesis models of
their spectral energy distribution (SED).  
The photometry used in these models encompasses the light from all components of a given galaxy,\footnote{Although this
may bias the SED fit if the two components are unrelated, the ground-based imaging that provides the rest-UV photometry is neither deep
enough nor sufficiently high resolution to reliably determine magnitudes for individual components visible in the {\it HST} imaging data.
This limitation likewise precludes us from using photometric redshift estimates to identify probable false pairs \citep[e.g.,][]{kartaltepe07,bundy09}.}
and $H_{160}$ magnitudes have 
been corrected for nebular line emission \citep[see details in][]{law12a}.
To fit the galaxy SEDs we use Charlot \& Bruzual (CB13) models
in combination with
a \citet{chabrier03} initial mass function and a constant star formation history.

\section{Spectroscopic Observations}
\label{nirspec.sec}

We obtained spectroscopic observations of these six galaxies
using the Keck/NIRSPEC longslit spectrograph \citep{mclean98}
with the slit rotated to lie along the position angle separating the two components visible in the {\it HST} imaging.  
Operating in low resolution mode, the NIRSPEC slit measures $42 \times 0.76$ arcsec, with a spectral resolution $R \sim 1400$ as measured from the widths
of skylines,
and a detector scale measuring 
0.143 arcsec pixel$^{-1}$ along the slit and 4.2 \AA\ pixel$^{-1}$ in the dispersion direction.
Integration times were typically 1 hour per target (see Table \ref{results.tab}), composed of four 15-minute exposures between which the target galaxy was
dithered along the slit.  Each target was acquired by blind offsets from a nearby reference star using astrometry derived from the {\it HST} imaging data.

Five of the six targets (Table \ref{results.tab}) were observed on 14 June 2012 in nearly photometric conditions
with near-IR $K$-band seeing $\sim 0.5$ arcsec FWHM.  In addition, one target (Q2343-BX429) was included from previous observations taken
in September 2003 under similar observing conditions that serendipitously had the long slit aligned with 
the pair separation axis to within $19^{\circ}$ (sufficient to incorporate both pieces
of the system).

We reduced the spectroscopic data using a hybrid scheme described by \citet{kulas12} that includes
cosmic ray rejection, image rectification, and a
two-dimensional sky subtraction algorithm (G. Becker, priv. comm)
to model the bright OH night-sky emission lines.
The resulting sky-subtracted spectra are stacked, wavelength calibrated to the heliocentric vacuum rest-frame
using the OH emission line features, and flux calibrated using observations of 
the A0 infrared standard stars HD 1160, HD 203856, and HD 18881 (Vega magnitudes $K = 7.04$, 6.84, and 7.14 respectively).
Two-dimensional reduced spectra (i.e., slit direction along the abscissa and spectral direction along the ordinate) 
for each of the target galaxies are shown in Figures \ref{mainfig1.fig} - \ref{mainfig4.fig}.
Except where noted otherwise, we  extract spectra of the primary, secondary, and (where applicable) tertiary 
objects using a 1 arcsec (7 pixel) wide box, and construct
a noise spectrum for each target by measuring the rms variations in a similar box along blank regions of the slit.

\section{Results}
\label{results.sec}

\begin{deluxetable*}{lccccccc}
\tablecolumns{8}
\tablewidth{500pt}
\tablecaption{Galaxy Emission Properties}
\tablehead{
\colhead{Galaxy} & \colhead{$z_{\rm H\alpha}$} & \colhead{$\lambda_{\rm H \alpha}$} & \colhead{Primary flux} & \colhead{Primary $\sigma_{\rm v}$} & \colhead{Secondary (Tertiary) flux\tablenotemark{a}} & \colhead{SLITPA\tablenotemark{c}} & \colhead{$N_{\rm exp}$ \tablenotemark{d}} \\
 & & (\AA) & ($10^{-17}$ erg s$^{-1}$ cm$^{-2}$) & (km $s^{-1}$) & ($10^{-17}$ erg s$^{-1}$ cm$^{-2}$) & 
}
\startdata
Q1217-BX116 & 2.1940 & $20967.5 \pm 0.2$ & $3.3 \pm 0.1$ & $< 91$ & $< 0.4$ & $135^{\circ}$ & 4\\
Q1217-MD16 & 2.6154 & $23733.6 \pm 0.8$ & $13.2 \pm 0.8$ & $103 \pm 13$ & $<1.8$ &  $184^{\circ}$ & 4\\
Q1623-BX543 & 2.5198 & $23105.8 \pm 0.7$ & $26.8 \pm 0.9$ & $180 \pm 8$ & $<1.8$ ($4.4\pm0.9$) & $205^{\circ}$ & 3\\
Q1700-MD103 & 2.3149 & $21760.7 \pm 0.5$ & $7.5 \pm 0.2$ & $102 \pm 12$ & $<0.5$ & $95^{\circ}$ & 5\\
Q2206-BM64 & 2.1948 & $20972.4 \pm 0.1$ & $4.8 \pm 0.2$ & $64 \pm 4$ & $<0.4$ ($<0.4$) & $64^{\circ}$ & 5\\
Q2343-BX429 & 2.1750 & $20842.4 \pm 0.2$ & $5.0 \pm 0.2$ & $66 \pm 7$ & $1.0 \pm 0.2$ & $24^{\circ}$ & 4\\
\enddata
\tablenotetext{a}{Flux limits are 3$\sigma$ limits for a spectroscopically unresolved feature at the wavelength of H$\alpha$ emission for the primary object.  Tertiary fluxes or flux limits are only given where applicable.}
\tablenotetext{b}{Redshift is from rest-UV absorption line spectrum.}
\tablenotetext{c}{Degrees East of North}
\tablenotetext{d}{Exposure time is $N \times 900$ seconds.}
\label{results.tab}
\end{deluxetable*}

\begin{figure*}
\plotone{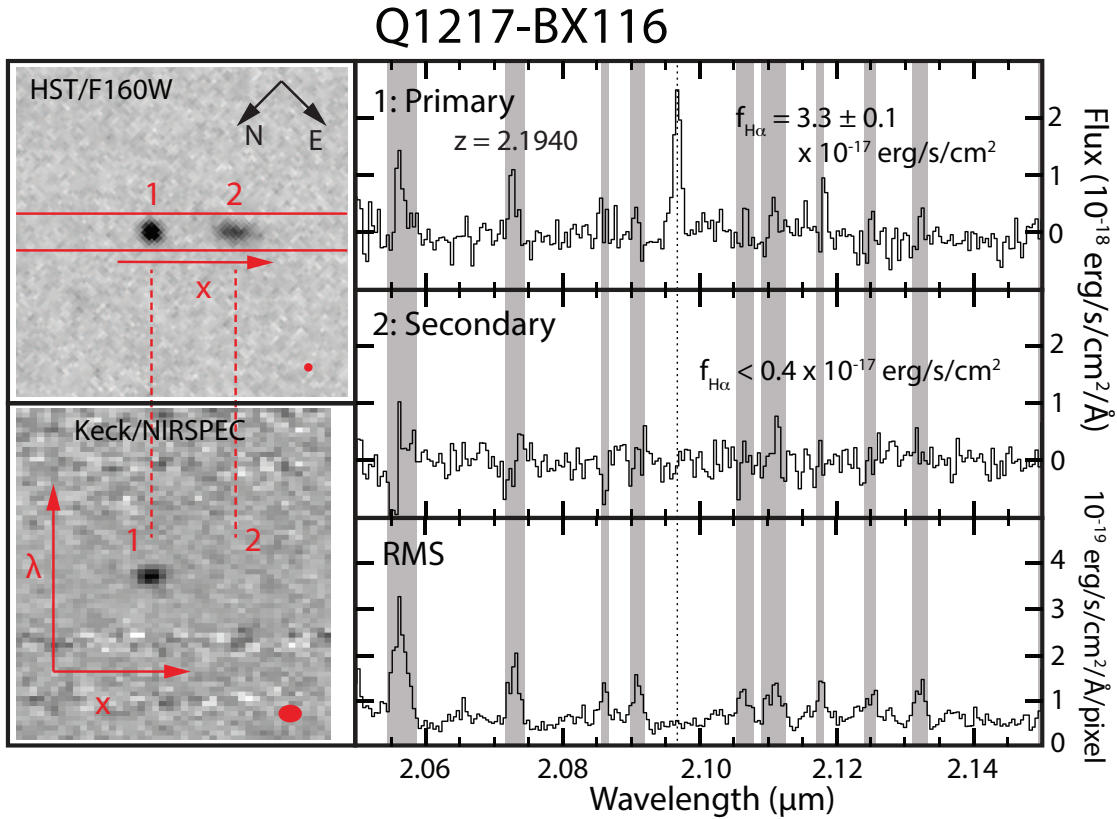}
\caption{Top left panel: {\it HST}/F160W imaging data for Q1217-BX116 ($z = 2.19$);  panel measures $7 \times 7$ arcsec$^2$. The color map has been inverted and uses an arcsinh stretch with the black point set to 28.46 AB pixel$^{-1}$ (23.0 AB arcsec$^{-2}$).
The orientation of the NIRSPEC slit is indicated by parallel red lines.  Bottom left panel: cutout of Keck/NIRSPEC 2d spectrum showing the corresponding spatial region along the slit
and $\pm$2000 km s$^{-1}$  in the wavelength dimension.  The colormap has been inverted and uses a linear stretch.  Individual components are numbered in both panels, and the red circle in each panel indicates the FWHM of the observational PSF.  Right-hand panels: spectra of the primary and secondary objects along with the associated RMS noise spectrum; 
grey shaded bars indicate the wavelengths of strong night-sky OH emission features.}
\label{mainfig1.fig}
\end{figure*}

\begin{figure*}
\plotone{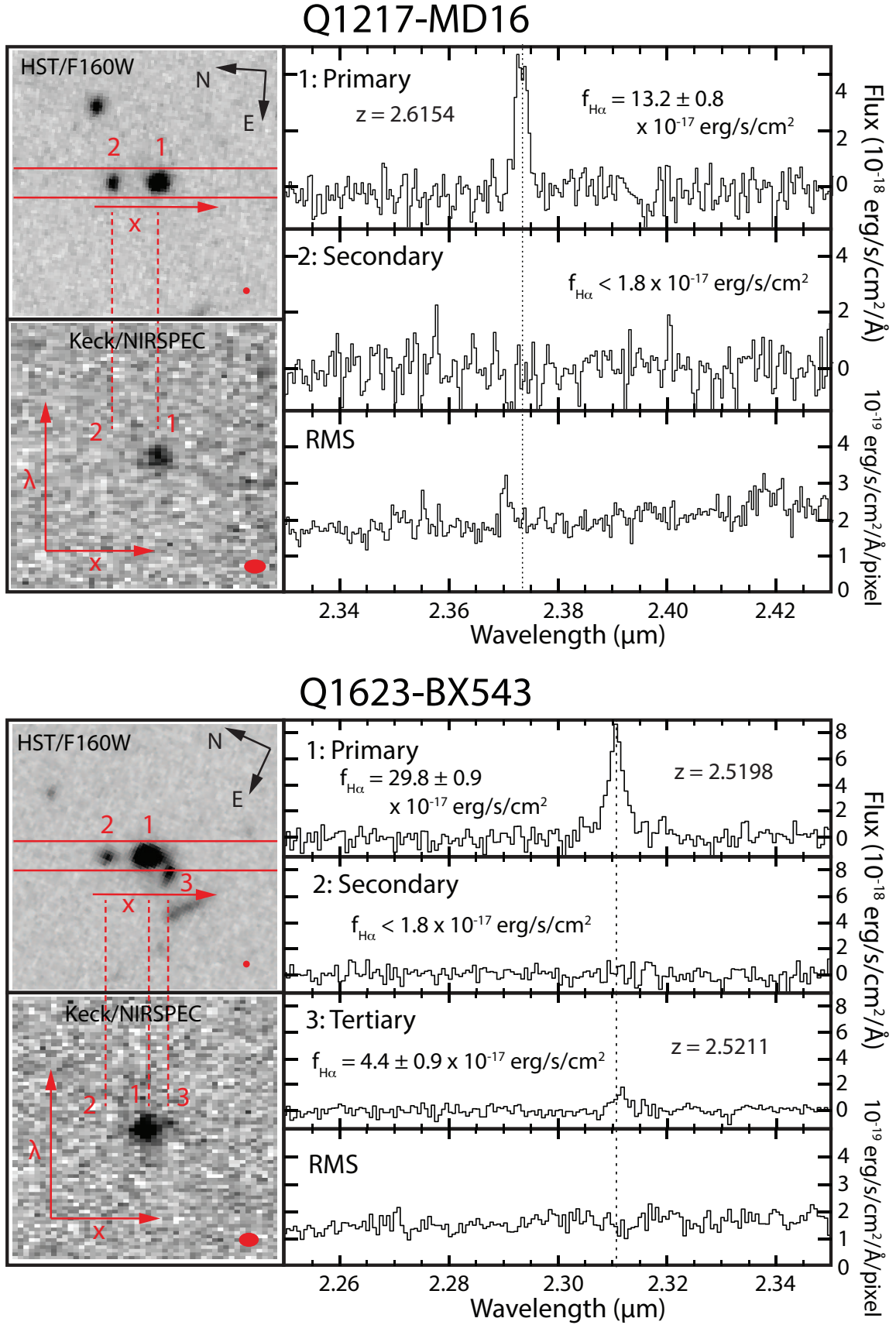}
\caption{As Figure \ref{mainfig1.fig}, but for Q1217-MD16 and Q1623-BX543 ($z = 2.62$ and $2.52$ respectively).  The NIRSPEC spectra for Q1623-BX543 show the  tertiary component, visible in the lower left-hand panel as a small spoke extending to the top right of the primary component.}
\label{mainfig2.fig}
\end{figure*}

\begin{figure*}
\plotone{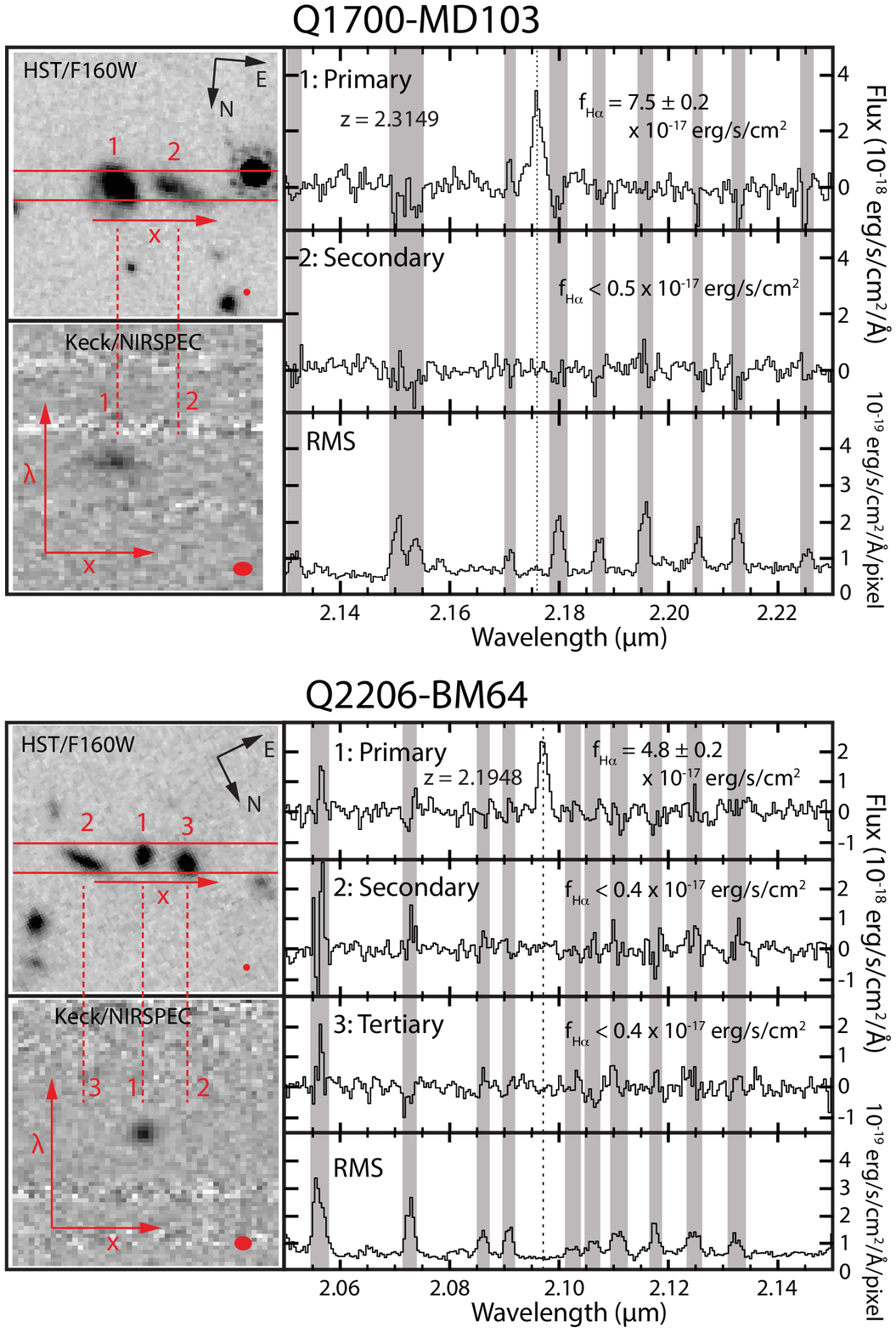}
\caption{As Figure \ref{mainfig1.fig}, but for Q1700-MD103 and Q2206-BM64 ($z = 2.31$ and $2.19$ respectively).  The object on the Eastern end of the Q1700-MD103 slit is a foreground star.}
\label{mainfig3.fig}
\end{figure*}

\begin{figure*}
\plotone{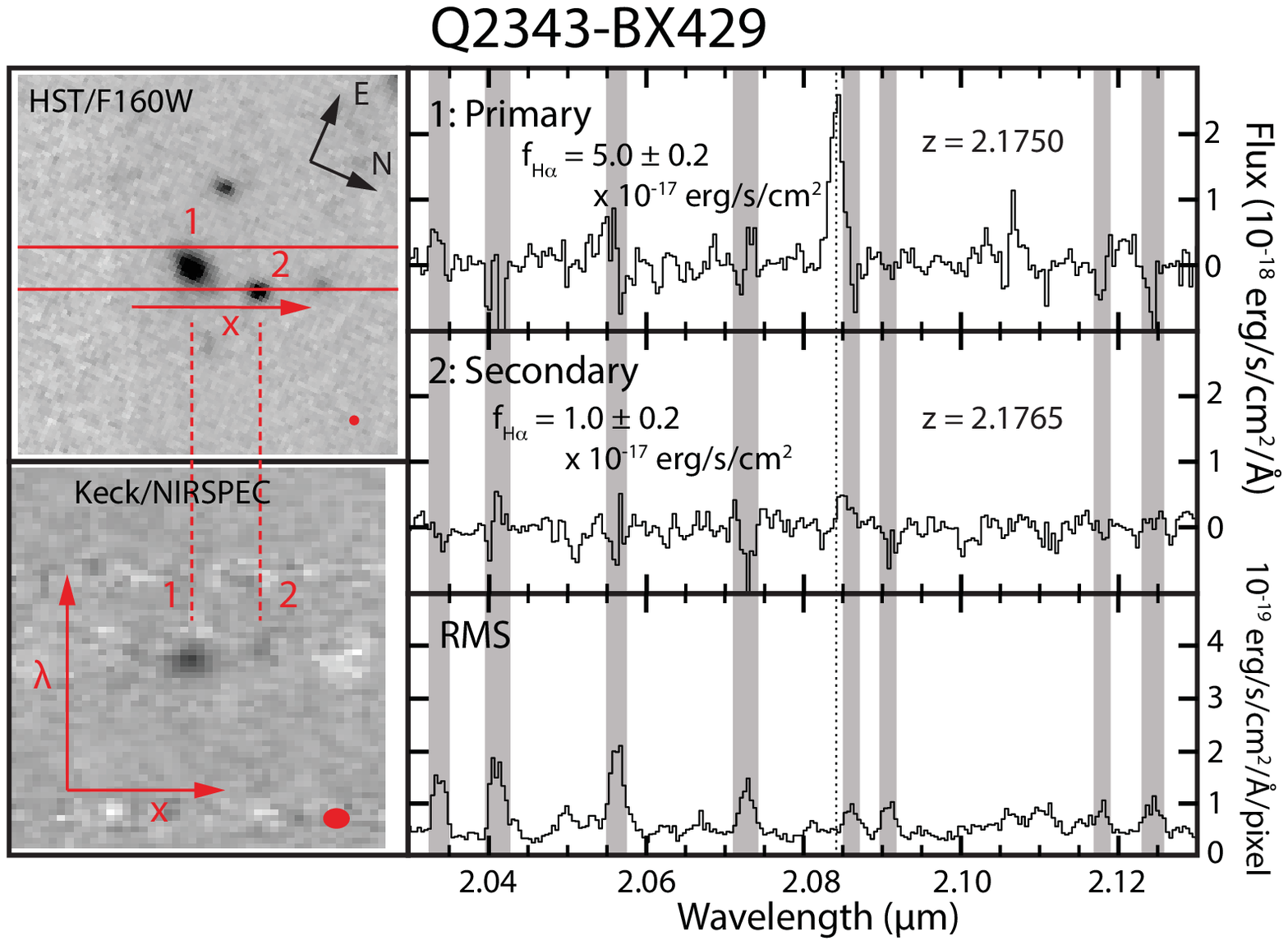}
\caption{As Figure \ref{mainfig1.fig}, but for Q2343-BX429 ($z = 2.17$).}
\label{mainfig4.fig}
\end{figure*}

We show the {\it HST}/F160W morphologies and NIRSPEC  spectra for our six target galaxies in Figures \ref{mainfig1.fig} --- \ref{mainfig4.fig},
along with estimates of the rms noise spectrum for each galaxy.  The sensitivity of these spectra is a function of
wavelength; using the rms spectra constructed from blank sky regions for each galaxy we estimate that the 
$3\sigma$ limit on spatially and spectroscopically unresolved emission ranges from $4 \times 10^{-18}$ \cgs\ at $\lambda = 2.1$ \micron\ to
$\sim 2 \times 10^{-17}$ \cgs\ at $\lambda = 2.4$ \micron\ where the thermal background of the Keck/NIRSPEC system becomes large.
We estimate the uncertainty in central wavelength and FWHM of individual lines from 
bootstrapped Monte Carlo tests in which realizations of the noise defined by the RMS spectra have been added to the object spectra.
We compute the velocity dispersion $\sigma_v$ of each target by subtracting off the instrumental resolution in quadrature from the measured FWHM of the H$\alpha$ emission line,
and estimate the spatial effective radius of nebular line emission along the slit by fitting a Gaussian profile,
subtracting the 0\farcs5 seeing in quadrature from the measured Gaussian FWHM, and dividing by 2.36 to obtain the effective $1\sigma$ radius.

\subsection{Q1217-BX116}
\label{bx116.sec}

Q1217-BX116 appears to be a relatively isolated double system in which both components have similar $H$-band magnitudes
($H_{160} = 24.56$ for the primary component, and 24.74 for the secondary component)
and are separated by 1\farcs7 in projection (corresponding to 14 kpc at $z = 2.19$).
The secondary component has a circularized effective radius $r_{\rm e} = 1.4$ kpc, significantly more elongated than the primary
($r_{\rm e} < 0.6$ kpc).\footnote{Circularized effective radii are derived using GALFIT \citep{peng02} and a model of
the {\it HST}/WFC3 PSF, and converted to kpc assuming that both pieces lie at the systemic redshift of the galaxy.
As detailed in \cite{law12a}, radii are likely systematically underestimated for sources with $H_{160} > 24$ AB.}
Based on SED modeling,
Q1217-BX116 is young (202 Myr) and low-mass ($M_{\ast} = 2.4 \times 10^9 M_{\odot}$) with a SFR of $11 \, M_{\odot}$ yr$^{-1}$.
Previous rest-UV spectroscopy obtained with Keck/LRIS \citep[see, e.g.,][and references therein]{steidel10} serendipitously included both pieces of the galaxy
and  showed extremely strong \lya\ emission with no clear absorption features.

The primary component of the pair is detected in \Ha\ to high confidence in our longslit NIRSPEC spectroscopy 
with an observed flux of $3.3 \pm 0.1 \times 10^{-17}$ \cgs.  As indicated by Figure \ref{mainfig1.fig} (lower left panel), there is no evidence of a tilt in the \Ha\ emission
profile, which has a roughly Gaussian profile along the spatial direction with a FWHM 0\farcs44 (i.e., consistent with an unresolved point source with intrinsic effective radius $r_{\Ha} < 1.2$ kpc given the $\sim$ 0\farcs5 seeing) and is spectrally unresolved with $\sigma_v < 91$ \kms.
Despite the strong detection of the primary in the middle of a relatively OH-free spectral region $\sim 700$ km s$^{-1}$ wide,
there is no evidence
of \Ha\ emission from the secondary component.  If the secondary component is not an unrelated interloper along the line of sight to Q1217-BX116,
we can use the RMS noise spectrum to
place a $3\sigma$ upper limit of $4 \times 10^{-18}$ \cgs\ on the \Ha\ emission flux from 
a spatially and spectroscopically unresolved source (i.e., roughly 10 times fainter than the primary).

\subsection{Q1217-MD16}
\label{md16.sec}

The two morphological components of Q1217-MD16 differ by a factor of 3.8 in brightness ($H_{160} = 23.44$ and 24.89 for the 
primary/secondary components respectively)
and are separated by 1\farcs1 (9 kpc at $z = 2.62$).   The primary and secondary components have similar circularized effective radii of
$r_{\rm e} = 0.6$ and 0.7 kpc respectively.  Similar to Q1217-BX116, SED models indicate that Q1217-MD16 is relatively young (80 Myr) and low-mass ($M_{\ast} = 7 \times 10^9
M_{\odot}$) with a SFR of $88 \, M_{\odot}$ yr$^{-1}$.
Previous Keck/LRIS rest-UV spectroscopy obtained at three different position angles suggested two components to the interstellar absorption line
features at $z=2.609$ and $z=2.616$, along with Ly$\alpha$ emission at $z=2.624$.

As indicated by Figure \ref{mainfig2.fig}, the primary component is well-detected in \Ha\ at $z=2.6154$ with flux $13.2 \pm 0.8 \times 10^{-17}$ \cgs,
consistent with recent Keck/MOSFIRE spectroscopy (Steidel et al. in prep) which detected \Hb\ emission at a redshift of $z=2.6168$.  
Although the nebular emission  is spatially unresolved
($r_{\Ha} < 1.2$ kpc after accounting for the observational seeing) it has a velocity dispersion of $\sigma_v = 103 \pm 13$ \kms.
Despite the suggestion of a two-component system from the rest-UV spectroscopy, no \Ha\ emission is detected at the location of the secondary morphological feature 
to a $3\sigma$ limit of $1.8 \pm 10^{-17}$ \cgs.

\subsection{Q1623-BX543}
\label{bx543.sec}

Q1623-BX543 is a complicated system with three distinct morphological components within $\sim 1$ arcsecond of each other (Figure \ref{mainfig2.fig})
and magnitudes $H_{160} = 22.82$, 25.12, and 24.48 for the primary, secondary, and tertiary components respectively.  
The circularized effective radius of the primary component is 0.9 kpc,  comparable to the secondary and tertiary features.
SED models indicate that Q1623-BX543 is young (9  Myr), low-mass ($M_{\ast} = 5 \times 10^9 M_{\odot}$), and forming stars extremely rapidly
at a rate of $515 \, M_{\odot}$ yr$^{-1}$ (see discussion in \S \ref{disc.sec}).

Previous $H$-band observations of Q1623-BX543 with adaptive-optics assisted OSIRIS 
integral-field spectroscopy  \citep{law09} found that components 1 and 3 are physically associated, with \othree\ emission 
features offset from each other by 125 \kms\ in velocity and 6.7 kpc (0\farcs8) in projected separation.  
The present NIRSPEC observations indicate that the primary component has an effective \Ha\ radius of $\sim 1.2 \pm 0.3$ kpc\footnote{Uncertainty is dominated by the uncertainty
in the observational PSF.} and
a broad velocity dispersion $\sigma_v = 180 \pm 8$ \kms\ with no apparent rotation about a preferred
kinematic axis.  Given the lower spectral and spatial resolution of the NIRSPEC data, the measured size and kinematics are fairly
consistent with the OSIRIS \othree\ $\lambda 5007$ observations ($R \sim 3400$,
PSF FWHM $\sim$ 0\farcs15) which 
suggested that this component has an effective radius of $1.1 \pm 0.1$ kpc and a net velocity dispersion
of $\sigma_v = 153 \pm 7$ \kms.\footnote{These are estimates of the integrated line width from the composite spectrum of the galaxy; the OSIRIS data
demonstrate that the galaxy has a mean internal velocity dispersion of $\sigma_{\rm mean} = 139$ \kms\ (RMS $=$ 32 \kms), with a velocity shear of
$39 \pm 4$ \kms.}

Although the NIRSPEC longslit observations were not intended
to measure the properties of the tertiary component (the position angle of the slit was chosen to cover the primary and secondary components) we nonetheless 
detect the tertiary feature at 0\farcs5 projected distance along the slit.  Extracting the spectrum from this location gives a flux of $4.4 \pm 0.9 \times 10^{-17}$ \cgs\ 
which is offset from the primary by $120 \pm 30$ \kms, consistent with the 125 \kms\ derived from OSIRIS \othree\ observations.
The tertiary feature is too faint to obtain a reliable estimate of its velocity dispersion, but our derived value
($\sigma_v = 110_{-70}^{+40}$ \kms) is consistent with previous OSIRIS \othree\ estimates of $\sigma_{\rm mean} = 60 \pm 10$ \kms.
We note that the dynamical mass ratio (8:1) of these two components as derived from the OSIRIS data is consistent with the stellar mass ratio
estimated using $H_{160}$ magnitude (5:1; see Table \ref{morphs.tab}).

The secondary morphological component  located 1\farcs1 (9 kpc) in projection to the northeast of the primary was not covered by the previous OSIRIS IFU data.
While the successful re-detection of the tertiary component gives us confidence in the ability of the NIRSPEC observations to discern faint features down to
small angular separations,
as indicated by Figure \ref{mainfig2.fig} we detect no \Ha\ emission from the secondary component to a $3\sigma$ limit of $1.8 \times 10^{-17}$ \cgs.

\subsection{Q1700-MD103}
\label{md103.sec}

Q1700-MD103 is significantly larger than most of the other galaxies, with primary (secondary) magnitude $H_{160} = 22.51$ (23.33)
and circularized effective radius $r_{\rm e} = 3.0$ (2.7) kpc.  The secondary morphological component is highly elongated with a centroid located
1\farcs5 East of the primary in projection (13 kpc at $z=2.31$).  Given the large effective radius of the primary, it is unsurprising
that the best-fit SED for the galaxy corresponds to an old (1 Gyr) and massive ($M_{\ast} = 5 \times 10^{10} M_{\odot}$) stellar population with
an ongoing SFR of 49 $M_{\odot}$ yr$^{-1}$.

Q1700-MD103 has strong \Ha\ emission, with a flux from the primary component of $7.5 \pm 0.2 \times 10^{-17}$ \cgs.
While the galaxy is a good candidate for rotation given its high mass \citep[see, e.g.,][]{fs09,law09,newman13} no rotation is evident along the axis
traced by the NIRSPEC slit.  Rather, it has an instrumentally-deconvolved velocity dispersion $\sigma_v = 102 \pm 12$ \kms\ and a large \Ha\
profile of effective radius $2.5 \pm 0.2$ kpc that roughly matches its rest-optical continuum radius.

In contrast, there is no trace of \Ha\ emission from the secondary component to a $3\sigma$ limit of $5 \times 10^{-18}$ \cgs.
In this case, we have the benefit of {\it HST}/ACS F814W imaging (from program GO-10581, PI: A. E. Shapley)
tracing rest-frame $\sim 2500$ \AA\ emission at the redshift of the primary component to help us understand this negative result.
As shown in \citet[][see their Fig. 9]{law12a},
the second component is extremely red compared to the primary ($I_{814} - H_{160} = 2.7$ vs 1.3) and only barely detected in the F814W imaging.  Since the primary and secondary
objects have such different colors it is probable that they either lie at different redshifts or that the secondary is significantly dustier/older; both explanations are consistent
with the lack of detection of \Ha.

\subsection{Q2206-BM64}
\label{bm64.sec}
 
Q2206-BM64 is a morphological triple system with magnitudes $H_{160} = 24.03$, 23.67, and 23.89 for the primary, secondary, and tertiary components respectively,
primary-secondary separation of 1\farcs1 (9.7 kpc) and a primary-tertiary separation of 1\farcs5 (12.8 kpc).
While the primary and secondary components are relatively circular with effective radii $r_{\rm e} = 0.8$ and 1.4 kpc respectively, the tertiary component is
highly elongated with $r_{\rm e} = 1.3$ kpc.  Q2206-BM64 has an SED best fit with a relatively old
(806 Myr) and massive ($M_{\ast} = 3 \times 10^{10} M_{\odot}$) stellar population and $37 \, M_{\odot}$ yr$^{-1}$ of ongoing star formation.

The unusual configuration of all three clumps in a nearly straight line meant that it was possible to obtain NIRSPEC spectroscopy of all three components simultaneously.
Although all three have similar colors based on our marginally resolved multi-wavelength ground-based imaging,
as indicated in Figure \ref{mainfig4.fig} only the primary component is detected in \Ha\ emission with a flux of $4.8 \pm 0.2 \times 10^{-17}$ \cgs.
While the emission from the primary is spatially unresolved (effective radius $\lesssim 1.2$ kpc), it has a velocity dispersion of $\sigma_{\rm v} = 64 \pm 4$ \kms.

\subsection{Q2343-BX429}
\label{bx429.sec}

Q2343-BX429 is composed of two clumps of magnitude $H_{160} = 24.71$ and 26.83 for the primary and secondary pieces respectively, with a projected separation of 1\farcs2 (10 kpc at $z=2.17$).  The primary component has an effective radius $r_{\rm e} = 1.5$ kpc, while the secondary component is spatially unresolved with $r_{\rm e} \leq 0.6$ kpc.
Similarly to Q1700-MD103 and Q2206-BM64, 
the SED of Q2343-BX429 is most consistent with an old (1.1 Gyr) and relatively massive ($M_{\ast} = 2 \times 10^{10} M_{\odot}$) stellar
population, with a small ongoing SFR of 14 $M_{\odot}$ yr$^{-1}$.

By coincidence, Q2343-BX429 was observed in July 2003 using Keck/NIRSPEC with the same instrumental setup 
and similar $K$-band seeing (0\farcs5)
as the rest of our galaxy targets, and with a position angle
sufficiently close to the separation vector between the two morphological components that both fell within the spectroscopic slit.  
The original reduction and analysis of these data have been described by \citet{erb04,erb06}.
As illustrated in Figure \ref{mainfig4.fig}, both the primary and secondary components are detected in the longslit spectra.  The primary component has a measured
\Ha\ flux of $5.0 \pm 0.2 \times 10^{-17}$ \cgs\ with an effective radius $< 1.2$ kpc and a velocity dispersion $\sigma_{\rm v} = 66 \pm 7$ \kms.
Although the secondary component is faint (\Ha\ flux $1.0 \pm 0.2 \times 10^{-17}$ \cgs) it is centered on exactly the location expected from the broadband {\it HST}/F160W imaging
data and has a centroid offset from the primary by $140 \pm 30$ \kms.
While this secondary component can be detected, 
we note that its flux calibration has a systematic uncertainty (of roughly a factor of 2) compared to the primary component since it lies on the edge of
the spectrograph slit.  It is also too faint, and too close to an OH skyline to obtain a reliable estimate of the intrinsic velocity dispersion.

\section{Discussion}
\label{disc.sec}

On the whole, the \Ha-derived properties of the target galaxies match well with expectations based on {\it HST}/WFC3 imaging and broadband photometry.
As indicated by Table \ref{SFR.tab}, the \Ha\ and rest-optical continuum radii are consistent with each other to within $\sim 20$\%, indicating that the ongoing star formation
is spatially coincident with the stellar mass accumulated during past star formation episodes.  This result is limited by our inability to measure true two-dimensional profiles
from the longslit data, but is consistent with previous findings from studies of $z \sim 2$ star-forming galaxies using adaptive-optics assisted integral field
spectroscopy \citep[e.g.,][]{law09,fs11}.

Similarly, the SFR derived from \Ha\ emission line flux and global SED modeling are in reasonable agreement.  We estimate the
SFR from the \Ha\ luminosity using the relation from \citet{kennicutt94} combined with a \citet{chabrier03} initial mass function
and an extinction correction based on the \citet{calzetti00} law \citep[see discussion by][]{erb12,reddy12} and the SED-derived $E(B-V)$. 
As shown in Table \ref{SFR.tab}, extinction-corrected \Ha\
and SED-based star formation estimates generally agree to within a factor $\sim 2$ or better.  The sole exception (Q1623-BX543) has a
formal best-fit stellar population age of less than 50 Myr, which is known to significantly inflate estimates of
the SFR above other indicators \citep{reddy12}.

\begin{deluxetable*}{lccccccc}
\tablecolumns{8}
\tablewidth{400pt}
\tablecaption{Galaxy Properties}
\tablehead{
\colhead{Galaxy} & \colhead{$r_{\rm e}$\tablenotemark{a}} & \colhead{$r_{\Ha}$\tablenotemark{b}} & \colhead{$E(B-V)$} & \colhead{$M_{\ast}$} &
  \colhead{SFR$_{\Ha}$\tablenotemark{c}} & \colhead{SFR$_{\rm SED}$} & \colhead{SFR$_{\rm Comp}$\tablenotemark{d}}\\
 & (kpc) & (kpc) & & $(10^{9} M_{\odot})$ & ($M_{\odot}$ yr$^{-1}$) & ($M_{\odot}$ yr$^{-1}$) & ($M_{\odot}$ yr$^{-1}$)
}
\startdata
Q1217-BX116 & 1.4 & $<1.2$ & 0.17 & 2.4 & $9\pm1$ & 11 & $<1$\\
Q1217-MD16 & 0.6 & $<1.2$ & 0.21 & 7.0 & $62\pm4$ & 88 & $<9$\\
Q1623-BX543 & 0.9 & $1.2\pm0.3$ & 0.30 & 5.0 & $158\pm5$ & 515 & $23\pm5$\\
Q1700-MD103 & 3.0 & $2.5\pm0.2$ &0.28 & 50 & $34\pm1$ & 49 & $<2$\\
Q2206-BM64 & 0.8 & $<1.2$ & 0.21 & 30 & $15\pm1$ & 37 & $<1$\\
Q2343-BX429 & 1.5 & $<1.2$ & 0.18 & 20 & $14\pm1$ & 14 & $3\pm1$\\
\enddata
\label{SFR.tab}
\tablenotetext{a}{PSF-corrected circularized effective radius of primary component derived from {\it HST}/WFC3 broadband imaging.  See \citet{law12a} for discussion of typical uncertainties.}
\tablenotetext{b}{PSF-corrected Gaussian half-light radius of primary component derived from Keck/NIRSPEC \Ha\ profile along the spectroscopic slit.}
\tablenotetext{c}{Extinction-corrected estimate for the primary component.  Uncertainties represent statistical uncertainty in the \Ha\ flux measurement, and do not incorporate systematic uncertainty in the SFR prescription adopted.}
\tablenotetext{d}{Extinction-corrected estimate for the morphological secondary companion (tertiary source for Q1623-BX543).}
\end{deluxetable*}

\begin{figure}
\plotone{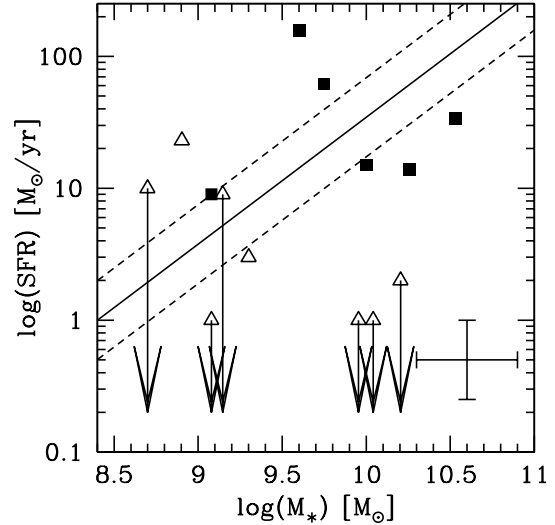}
\caption{Estimated stellar mass and SFR for the primary (filled squares) and secondary/tertiary (filled triangles) objects.  Stellar masses are estimated from the best-fit stellar population model of the integrated light of the system (divided up between components according to their relative $H_{160}$ flux), SFR are derived from the nebular emission line fluxes and upper limits.  The error bar in the lower right corner of the panel indicates the typical uncertainty of a given point (which is dominated by systematics).  The solid line represents the track of the star forming galaxy main sequence at $z = 1.5-2.5$ taken from \citet{wuyts11}, dashed lines indicate SFR a factor of two above and below this track.}
\label{ssfr.fig}
\end{figure}

Of the six galaxies observed, two  
were confirmed to have nearby companions
(Q1623-BX543, Q2343-BX429; mass ratios 5:1 and 7:1 respectively) 
whose projected separations and relative velocities indicate that they are likely
in the process of merging with the central galaxy.\footnote{Although as discussed by \citet{patton08} and \citet{lotz11},
even close pairs with similar spectroscopic redshifts are not {\it certain} to be merging systems since small redshift differences can 
correspond to large separations along the line of sight.}
As indicated by Figure \ref{ssfr.fig}, all six primary objects and both confirmed companions lie near the star forming galaxy main sequence for $z = 1.5 - 2.5$ galaxies \citep[e.g.,][]{wuyts11}.
This pair confirmation rate (1/3) is consistent with the expectation that $\sim$ 1/2 of apparent pairs
with angular separations $\sim$ 1-2 arcsec are physically unrelated superpositions along the line of sight based on the statistical distribution of sources
in the {\it HST}/WFC3 imaging fields \citep[see discussion by][]{law12a}.  This explanation is perhaps particularly likely for Q1700-MD103
and Q2206-BM64, for which the putative companions fall well below the star formation main sequence (lower-right open triangles in Fig. \ref{ssfr.fig}).
However, our results are also compatible with the hypothesis
that the SFR in merging pairs can differ
by a factor of 10 between each of the components, even when the components
have similar rest-optical continuum magnitudes.  Such an explanation was recently put forward by
\citet{schmidt13}, who used 3D-{\it HST} grism spectroscopy to show that apparent morphological pairs tended to have nebular emission-line indicators of star formation concentrated in just one component of the pair, ascribing the difference  to different gas content in the components of the merger.  Here
we push the SFR threshold required for detection an order of magnitude deeper than \citet{schmidt13}, but similarly find an absence of evidence
for pronounced star formation in the merging companions.

In previous analyses of the $z \sim 2$ star forming galaxy population we found 
\citep{law07a,law12a,law12c} that rest-frame UV and optical continuum morphology was largely
decoupled from other physical properties, with merger-like galaxies 
having a distribution of stellar masses and SFR statistically consistent with the non-merger population.  However, as discussed by \citet{law12a}
apparent morphological pairs tended to have marginally
higher SFR surface density than the rest of the $z\sim 2$ star forming galaxy population
 ($\langle \Sigma_{\rm SFR} \rangle$ = 10 $M_{\odot}$ yr$^{-1}$ kpc$^{-2}$ for mergers vs
 4 $M_{\odot}$ yr$^{-1}$ kpc$^{-2}$ for non-mergers, with 0.2\% confidence in the null hypothesis of being drawn from
 the same distribution).  Additionally, such pairs were much more likely to show \lya\ in emission than the rest of the galaxy sample, with many of the strongest
 \lya\ sources (e.g., Q1217-BX116) having a double-component morphology \citep{law12c}.  This latter trend was also noted
 by \citet{cooke10}, whose study of spectroscopically-confirmed Lyman Break Galaxy (LBG) pairs at $z \sim 3$ found that pairs were much more likely than 
 non-pairs to exhibit \lya\ in emission \citep[although c.f.][]{shibuya14}.

While it is impossible to draw statistically robust conclusions from our small pilot sample, it is nonetheless informative to consider
how the properties of our two spectroscopic pairs
compare to those of the overall star forming galaxy population from which they were drawn
\citep[$\langle M_{\odot} \rangle \sim 10^{10} M_{\odot}$, $\langle \textrm{SFR} \rangle \sim 30 \, M_{\odot}$ yr$^{-1}$, see Fig. 18 of][]{law12a}.
As detailed in \S \ref{bx543.sec} and \ref{bx429.sec} above, 
Q1623-BX543 has higher than average SFR and $\Sigma_{\rm SFR}$,
and lower than average stellar mass and circularized effective radius.  In contrast Q2343-BX429 has lower than average SFR,
higher than average stellar mass, and $r_{\rm e}$ and $\Sigma_{\rm SFR}$ fairly typical of the $z \sim 2$ star forming galaxy population.
Neither galaxy shows \lya\ in emission based on rest-frame UV spectra.
Although our findings are therefore marginally inconsistent with the findings of \citet{cooke10} and \citet{law12c} that merging pairs seem more likely
than non-pairs to show \lya\ emission, they generally match other findings \citep{law07a,law12a,lee13,schmidt13}
that spectroscopic pairs do not have physical properties any different
on average from those of the $z \sim 2$ star forming galaxy population as a whole.

Looking forward, we note that it will soon be possible to resolve such discrepancies.
Using catalogs of close-pair candidates extracted from large-area morphological surveys such as CANDELS, high-sensitivity
follow-up spectroscopy
using recently-commissioned multi-object NIR spectrographs such as MOSFIRE \citep[e.g.,][]{kriek14} and KMOS
\citep[e.g.,][]{wisnioski14} is likely to spectroscopically confirm samples of a few tens to hundreds of genuine mergers for only a modest
investment in observing time.
Such statistically large, representative 
samples will allow us to determine the influence of major merging events on the evolution of gas and stellar populations
in galaxies in the young universe.

\acknowledgements

These results are based 
in part on data obtained at the W. M. Keck Observatory, which is operated as a scientific partnership among the California Institute of Technology, the University of 
California, and NASA, and was made possible by the generous financial support of the W. M. Keck Foundation.
DRL and CCS have been supported by grant GO-11694 from the Space Telescope Science Institute,
which is operated by the Association of Universities for Research in Astronomy, Inc., for NASA, under contract NAS 5-26555.
CCS has been supported by the US National Science Foundation through grants AST-0606912 and AST-0908805.
AES acknowledges support from the David and Lucile Packard Foundation.
DRL thanks Kristen Kulas for assistance obtaining the Keck/NIRSPEC spectroscopy,
Dawn Erb for making available previous NIRSPEC observations of Q2343-BX429, 
and George Becker for sharing a copy of his IDL-based NIRSPEC data reduction code.
Finally, we extend thanks to those of Hawaiian ancestry on whose sacred mountain we are privileged to be guests.

\end{document}